\documentclass{webofc}
\usepackage[varg]{txfonts}   % Web of Conferences font
\usepackage{hyperref}
\newcommand{\muB}{\mu_\text{B}}
\newcommand{\nB}{n_\text{B}}
\newcommand{\mE}{m_\text{E}}
\newcommand{\Lh}{\Lambda_\text{h}}
\newcommand{\ud}{\mathrm{d}}
\newcommand{\II}[1]{I_\text{#1}}
\newcommand{\D}[1]{D^{#1}}
\newcommand{\DZ}{\Delta_0}
\newcommand{\Dz}[1]{\DZ^{#1}}
\newcommand{\DN}[1]{\Delta_{#1}}

\newcommand{\DDN}[1]{D_{#1}}

\renewcommand{\vec}{\mathbf}
\newcommand{\eq}{Eq.~}
\renewcommand{\Ref}{Ref.~}
\newcommand{\Refs}{Ref.~}
\newcommand{\trT}{{\rm tr}}
\newcommand{\tr}[1]{{\rm tr}\left[ #1 \right]}
\begin{document}
\title{Improving the cold quark-matter pressure via soft interactions at N3LO}
%
% subtitle is optionnal
%
%%%\subtitle{Do you have a subtitle?\\ If so, write it here}

\author{\firstname{Tyler} \lastname{Gorda}\inst{1}\fnsep\inst{2}\thanks{\email{tyler.gorda@physik.tu-darmstadt.de}}         % etc.
}

\institute{Technische Universität Darmstadt, Department of Physics, D–64289 Darmstadt, Germany
\and
Helmholtz Research Academy for FAIR, D–64289 Darmstadt, Germany
}

\abstract{%
    The propagation of long-wavelength gluons through a dense QCD medium at high baryon chemical potential $\muB$ is qualitatively modified by the effects of screening, arising from scatterings off the high-momentum quarks in the medium. This same screening phenomenon also impacts gluons occurring in loop corrections to the pressure of cold quark matter, leading to contributions from the parametric scale $\alpha_s^{1 / 2} \muB$, starting at next-to-next-to-leading order (N2LO) in the strong coupling constant $\alpha_s$. At next-to-next-to-next-to-leading order (N3LO), interactions between these long-wavelength gluonic modes contribute to the pressure. These interaction corrections have recently been computed in Refs.~\cite{Gorda:2021kme,Gorda:2021znl}, and the inclusion of these interactions slightly improves the convergence of the equation of state of cold quark matter. In these proceedings, we present these results and provide details summarizing how this lengthy calculation was performed.
}
\maketitle
\section{Introduction}
\label{intro}

In recent years, the cold (zero-temperature, $T = 0$) quark-matter (QM) equation of state (EoS) has been used as a nontrivial  high-density limit \cite{Komoltsev:2021jzg} to constrain the EoS of neutron-star (NS) matter at much lower baryon densities $\nB$ \cite{Kurkela:2014vha,Annala:2017llu,Most:2018hfd}. Typically, this cold-QM EoS can only be used at densities $\nB\gtrsim40n_0$, with $n_0 \approx 0.16$~fm$^{-3}$ corresponding to nuclear saturation density, where calculations in perturbative Quantum chromodynamics (pQCD) \cite{Kurkela:2009gj,Kurkela:2014vha,Gorda:2021znl,Gorda:2021kme} show a small renormalization-scale dependence and are thus under perturbative control. 

\begin{figure}[t]
% Use the relevant command for your figure-insertion program
% to insert the figure file.
\centering
\includegraphics[width=0.55\textwidth]{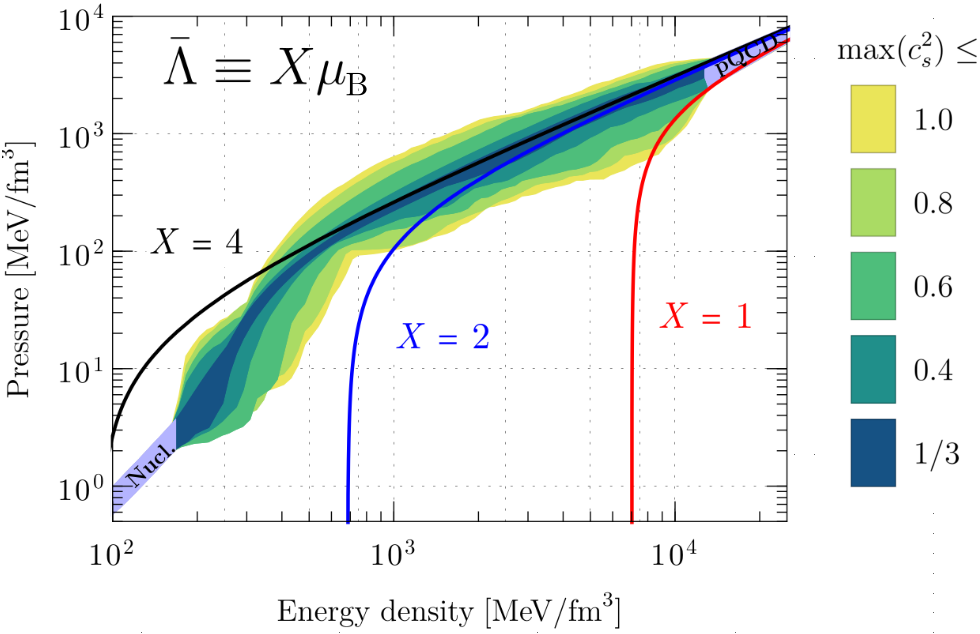}
    \caption{An illustration of the renormalization-scale dependence of the cold-QM EoS at N2LO \cite{Kurkela:2009gj}. The cold-QM EoS is shown for the three different choices of the renormalization parameter $X = 1, 2, 4$, plotted on top of the regions of viable NS-matter EoSs as determined in Ref.~\cite{Annala:2019puf} using robust theoretical and astrophysical constraints. These regions are color-coded by the maximal value of the speed of sound squared $c_s^2$ obtained at any point along the EoS, with lower $\max(c_s^2)$ regions plotted on top of higher ones.}
\label{fig:renorm_scale_ep}       % Give a unique label
\end{figure}

In calculations at high density, one typically chooses the renormalization scale $\bar{\Lambda}$ to be proportional to the baryon chemical potential 
\begin{equation}
    \bar{\Lambda} = X \muB,
\end{equation}
and furthermore chooses $X = 2$ as the central value and $X \in[1,4]$ as the reasonable range quantifying the uncertainty \cite{Kurkela:2009gj,Kurkela:2014vha}. As one sees from Fig.~\ref{fig:renorm_scale_ep}, the uncertainty on the cold QM EoS at next-to-next-to-leading order (N2LO) in the strong coupling $\alpha_s$ (computed in Ref.~\cite{Kurkela:2009gj}) quickly explodes below these densities where pQCD calculations show good convergence. Moreover, as of yet, no further prescription can be used to constrain the coefficient $X$, such as the principle of minimal sensitivity (PMS) \cite{Stevenson:1982qw}, as the current functional form of the EoS of cold QM does not exhibit a stationary point as a function of $X$. However, including higher-order next-to-next-to-next-to-leading order (N3LO) corrections to the cold QM EoS may change this, as new functional forms will be added to the pressure. If this does happen, it could lead to a drastic improvement in the behavior of this thermodynamic quantity.

Improving the EoS of cold QM from N2LO to N3LO is complicated by the physics of dynamical screening of long-wavelength, low-energy gluonic modes. This phenomena can be understood by examining the dispersion relation for gluons with frequency $\omega$ and momenta $\vec{k}$ within loop corrections in cold QM:
\begin{equation}
    - \omega^2 + |\vec{k}|^2 + \Pi(\omega, \vec{k}) = 0.
\end{equation}
Parametrically, $\Pi = O(\mE^2) = O(\alpha_s \muB)$, with $\mE$ a screening mass, related to the one-loop Debye mass.\footnote{For a single massless quark with quark chemical potential $\mu_q$, $\mE^2 \equiv (2 / \pi ) \alpha_s \mu_q^2$} This holds true even for $\omega, |\vec{k}| \ll \muB$, and so for gluons with $\omega, |\vec{k}| \simeq \mE$, the self-energy corrections are as large as the free term. For these modes, one cannot treat the self energy as a small perturbation, and instead one must use a resummed gluonic propagator, with the self energy included in the denominator.

At high temperatures, where $\omega$ can only take on discrete values given by the Matsubara modes $\omega_n \equiv 2 \pi n T$, $n \in \mathbb{Z}$, only the $n = 0$ mode has low enough energy to require resummation, which can be treated within the dimensionally reduced effective field theory (EFT) of electrostatic QCD (EQCD) \cite{Appelquist:1981vg,Kajantie:1995dw,Braaten:1995cm}. However, within cold QM at $T = 0$, the discrete Matsubara summation becomes replaced by a normal integral over Euclidean frequencies, and thus there is no longer a clear separation between those energies which are small enough to require resummation and those which are not. 
This in turn means that there is no clear way to separate the effects from the long-wavelength screened gluons as a distinct EFT within cold QM, and rather the pressure will receive contributions of three different types:
\begin{enumerate}
    \item \emph{hard} contributions, from short-wavelength modes with $\omega, |\vec{k}| \gtrsim \muB$ and their interactions, 
    \item \emph{soft} contributions, from long-wavelength modes with  $\omega, |\vec{k}| \lesssim \mE$ and their interactions, 
    \item \emph{mixed} contributions, sensitive to interactions between the hard and soft modes.
\end{enumerate}
These contributions to the pressure $p$ can be organized as follows \cite{Gorda:2021kme}:
\begin{align} 
    \nonumber
    p = p_{0} +  \alpha_s p_1^h &+ \alpha_s^2 p_2^h + \alpha_s^3 p_3^h \quad (\text{hard modes})\\ 
    \nonumber
        &+ \alpha_s^2 p_2^s + \alpha_s^3 p_3^s \quad (\text{soft modes}) \\ 
             &\phantom{+ p_2^s \alpha_s^2 \,\,}+\alpha_s^3 p_3^m \quad (\text{both}).
\label{eq:p_decompose}
\end{align}
The above equation contains all contributions up to N3LO. That the soft modes do not enter until N2LO follows from the phase-space suppression required from the soft kinematics: ${\rm d}^4 P = O(\alpha_s^2 \muB^4)$ for soft modes. That the soft interactions remain perturbative, with interaction corrections leading to additional powers of the coupling (unlike at high $T$---see \cite{Ghiglieri:2020dpq} for a discussion) follows from a more subtle argument discussed in Refs.~\cite{Gorda:2018gpy,Gorda:2021kme}, but the physical reason is simple: at $T = 0$, gluons are not thermally populated by the medium, and thus do not receive an infrared (IR) over-occupation (Bose enhancement) as they do at high $T$. The correct treatment of the interacting soft scale is given by the hard-thermal-loop (HTL) [or hard Dense loop (HDL)] effective theory \cite{Braaten:1989mz,Manuel:1995td}.

The lack of a clear split between the hard and soft modes leads to an ambiguous \emph{semisoft} range of energies and momenta $\mE \ll \omega, |\vec{k}| \ll \muB$ in which resummations are not necessary, but one may still use the kinematic simplifications of the HTL theory. This leads to IR divergences in the hard sector which will cancel against ultraviolet (UV) divergences in the soft sector (as well as mixed IR-UV divergences in the mixed sector). These canceling divergences also leads to the $p_i^{s / m / h}$ depending on a factorization mass scale $\Lh$, which cancels out when summing over the kinematic sectors at a given order. When performing calculations in dimensional regularization, as done in Refs.~\cite{Gorda:2021kme,Gorda:2021znl}, this $\Lh$ appears in the integration measure, as an additional $\overline{\text{MS}}$ scale.

\section{Detailed summary of the N3LO soft calculation}
\label{sec:calc_details}

\begin{figure}[t]
% Use the relevant command for your figure-insertion program
% to insert the figure file.
\centering
\includegraphics[width=0.68\textwidth]{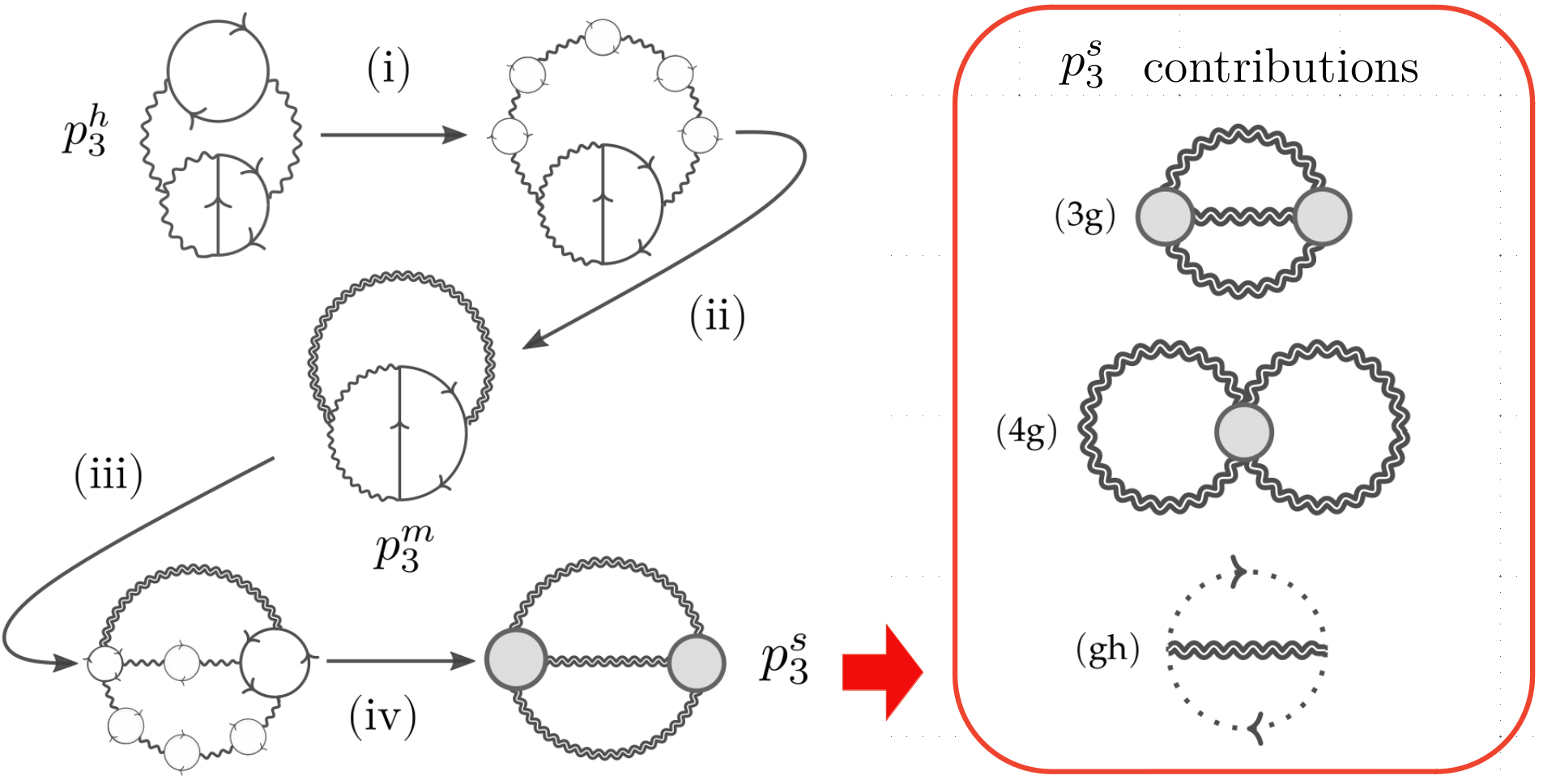}
    \caption{(Left) An illustration of how to arrive at contributions from different momentum scales from a hard diagram. When a gluon momentum in a four-loop diagram becomes soft (i), additional self-energy corrections do not change the order of the diagram. (ii) This soft gluon line can be resummed within the HTL theory, producing the HTL propagator denoted by a thick wavy line and leads to a three-loop `mixed' contribution if the other gluon momenta are hard. (iii) If the other gluon momenta in the diagram also become soft, then the other gluon propagators and interaction vertices must be dressed with additional loops as well, giving rise to (iv) a fully soft HTL contribution to $p_3^s$. (Right) All the two-loop HTL diagrams that contribute to $p_3^s$, derivable in this manner.} 
\label{fig:zorro}  
\end{figure}

At N3LO, the $p_3^s$ contribution in Eq.~\eqref{eq:p_decompose} arises from two-loop HTL diagrams. This can be seen following the analysis in Fig.~\ref{fig:zorro}. Note that fermion lines never need to be resummed at high density because the soft fermion modes are Pauli blocked by the medium. The gray blobs at the vertices in this figure indicate the use of one-loop HTL-corrected vertex functions, which are a sum of a bare ($\Gamma_0$) and HTL ($\delta \Gamma$) vertices: $\Gamma\equiv\Gamma_0+\delta \Gamma$. The thick gluon lines correspond to propagators dressed with one-loop HTL self-energies ($\Pi$). (We are here suppressing Lorentz structure). In Ref.~\cite{Gorda:2021kme}, the calculation of these diagrams was performed for massless quarks in dimensional regularization using the $\overline{\text{MS}}$ scheme, taking  $d = 3 - 2\epsilon$, and was performed in Feynman ($\xi = 1$) gauge, though it was checked that the full result is gauge invariant, as also stated in Ref.~\cite{Andersen:2002ey}. Note that this reference also studied these two-loop diagrams at high temperatures, but notably did so by expanding in the ratio $\mE / T$. At $T = 0$, such an expansion is not valid and the entire resummed diagrams must be computed. Below, we provide a summary of key details from Ref.~\cite{Gorda:2021kme}, in which the full, unexpanded calculation is performed at $T = 0$.

These three HTL graphs shown in Fig.~\ref{fig:zorro} contribute to the pressure according to
\begin{equation}
    \alpha_s^3 p_3^s = g^{2} N_c d_A 
    \bigl[\II{3g} + \II{4g} +  \II{gh}  \bigr], 
   % \equiv g^{2} N_c d_A I_\text{tot},\;\;\; 
\label{eq:all} 
\end{equation}
with the individual expressions given in Ref.~\cite{Gorda:2021kme}. Here, $g \equiv \sqrt{4 \pi \alpha_s}$, $N_c$ is the number of quark colors, and $d_A \equiv N^2_c - 1$ is the number of gluons. One important property of the HTL gluon self energy $\Pi$ is that it does not depend on the magnitude of the (Euclidean) four momentum of the gluon. This means that each loop integral in the diagram is divergent in the UV, leading to each of the above HTL graphs contributing terms to the pressure at orders $1/\epsilon^2$, $1/\epsilon$, and $\epsilon^0$. Their sum can thus be expressed in the form 
\begin{equation}
        \alpha_s^3 p_3^s \!= 
        \frac{g^{2} N_c d_A \mE^{4}}{(2\pi)^{6}}  \!
        \left( \frac{\mE}{\Lh} \right)^{\!\!-4\epsilon} \!
        \left[ \frac{p_{-2}}{(2 \epsilon)^{2}} + \frac{p_{-1}}{2 \epsilon} + p_{0} \right], 
\end{equation}
and from the detailed computation \cite{Gorda:2021kme}, one finds the following compact, final expressions:
\begin{align}
    \label{eq:pm2_result}
    p_{-2} ={}&  
        \frac{11}{6} 
        \int_{\Omega} \trT{\left[ \hat{\Pi^{2}}(\hat{K})\right]} 
        = 
        \frac{11 \pi^{2}}{24},\\
    \label{eq:pm1_result}
    p_{-1} ={}&    
        \int_{\Omega}
        \Biggl\{
             \frac{19+ 11\pi^2}{72}
           % + \frac{11}{72} \pi^{2}
            - \frac{11}{6} 
            \trT{\left[
               \hat \Pi^{2}( \hat{K}) \ln \bigl[ \hat \Pi(\hat{K}) \bigr]   \right]          
            }
        %\nonumber \\
       +\int_{0}^{\pi/2} \frac{\ud \chi \sin(2 \chi)\,\trT{\left[ \delta  \hat \Gamma^2_{3\text{g}}\bigl(\hat{K} \sin\chi, \hat{P} \cos \chi\bigr) \right]}}{24 [1+\sin(2\chi)  \hat K \cdot \hat P]} 
        \Biggl\} \nonumber \\
%      &=&11.6840(15) \\
    \approx{}& 8.7432  + 2.5068 + 0.4340(15) = 11.6840(15), \\
    \label{eq:p0_result}
    p_0 ={}& 17.150(7),
\end{align}
where the hatted notation on the HTL functions indicates that the $\mE$ has been scaled out, and the trace-and-square notation indicates that the Lorentz indices of the two corresponding HTL functions are fully contracted with one another. The hatted notation on the four-momenta indicates that their magnitudes have been divided out, leaving a unit Euclidean four-vector. The angular integral $\int_{\Omega}$ and the angle $\chi$ will be defined further below.  Let us now gives some details from Ref.~\cite{Gorda:2021kme} of how this result is derived. 

Within the full expression, the UV-sensitive and UV-finite terms are not clearly separated, since the resummed gluon propagators are a sum of an infinite number of terms, each scaling differently in the UV.  Hence, a strategic approach is required to extract the $p_{-2}$ and $p_{-1}$ coefficients from the resummed two-loop integrals. As an observation, we note the following: the logarithmically UV-sensitive terms (those which do not simply vanish in dimensional regularization) are related to pieces of the full expression that scale as $\mE^{4}$ times a dimensionless integral in the UV, as such a structure allows it to contribute throughout the whole UV tail of the integral. If one could isolate all such terms, while keeping their resummed nature intact, then one would be able to extract the UV divergences clearly.

This motivates the following notation: 
\begin{align}
    \DN{n}^{\mu \nu}(K) \equiv (-1)^{n} \frac{[\Pi(K)^{n}]^{\mu \alpha}}{(K^{2})^{n}}\Dz{\alpha \nu}(K); \qquad
    \DDN{n}^{\mu \nu}(K) \equiv{} (-1)^{n} \frac{[\Pi(K)^{n}]^{\mu \alpha}}{(K^{2})^{n}} \D{\alpha \nu}(K),
\end{align}
where $\DZ$ is the bare and $D$ is the HTL-resummed gluon propagator, and $(\Pi^{n})^{\mu \alpha} \equiv \Pi^{\mu \nu_1}\Pi^{\nu_1 \nu_2} \cdots \Pi^{\nu_{n-1} \alpha}$. The first of these definitions corresponds precisely to the terms occurring in the UV expansion of the HTL propagator. Together, these allow one to peel off bare parts of the resummed propagator without expanding: the relation $D=\sum_{k=0}^{n-1}\DN{k}+\DDN{n} $ holds exactly for any $n \ge 1$, and both the bare and resummed pieces scale like $\DN{n}(K) \sim \DDN{n}(K) \sim \mE^{2n} K^{-2(n+1)}$ in the UV. If one chooses an $n$ large enough ($n = 2$ suffices in all cases here), one can substitute this exact relation for all of the resummed propagators in each contribution $\II{3g}$, $\II{4g}$, $\II{gh}$; multiply out the terms; and select those scaling as $\mE^4$ in the UV.\footnote{The propagator manipulations also introduce scaleless integrals, containing only $\DN{n}$ terms, which vanish in dimensional regularization. These terms additionally have too few powers of $\mE$ to contribute at N3LO.} To do this, one must also take note of the UV scaling of the vertices $\Gamma = \Gamma_0 + \delta \Gamma$: for the three-gluon vertex, $\Gamma_{0,3\text{g}} \sim K$ and $\delta \Gamma_{3\text{g}} \sim \mE^{2} K^{-1}$,
%in the UV, 
while for the four-gluon vertex, $\Gamma_{0,4\text{g}} \sim K^{0}$ and $\delta \Gamma_{4\text{g}} \sim \mE^{2} K^{-2}$.%
%in the UV.
\footnote{Note that there are multiple momentum scales for each vertex.  These scalings hold for each momentum in the UV.} %Using these scalings and propagator machinery, we can clearly divide each contribution $\II{3g}$, $\II{4g}$, and $\II{gh}$ into multiple logarithmically UV-sensitive and UV-finite pieces.  
We shall now discuss in detail the manipulations that are performed on the UV-sensitive terms in Ref.~\cite{Gorda:2021kme}, leaving a discussion of the finite terms to the end. 

In the UV-sensitive terms, one can perform the contractions and arrive at compact expressions, since these terms often involve at least one bare propagator, which directly contracts the vertices. Via these direct contractions, one can make extensive use of the generalized Ward identities for the HTL vertex functions and reduce some four-point functions to three-point functions and some three-point functions to (two-point) self energies. Combining all the UV-sensitive terms, one arrives at the following elegant expression for the UV-sensitive part of $\II{tot} \equiv \II{3g} + \II{4g} + \II{gh}$: 
\begin{align}
\label{eq:ItotUV}
    \bigl[\II{tot}\bigr]^{\text{UV}} =& \int_{KP} \biggl\{
        \frac{2P^{2}}{R^{2}}
        [\hat{P} \cdot \DDN{1}(K) \cdot \hat{P}] 
        \tr{\DDN{1}(P)} -\frac{1}{4} 
        \tr{\DDN{1}(K)} 
        \tr{\DDN{1}(P)} 
        +
        \frac{2 K^{2}}{P^{2} R^{2}}
        \tr{\DDN{2}(K)} \nonumber \\
        &+
        \frac{d-1}{R^{2}}
        \bigl[
            \hat{P} \cdot \DDN{2}(K) \cdot \hat{P}
        \bigl] 
        +
        \frac{1}{12} 
        \delta \Gamma^{\mu \nu \rho}_{3\text{g}}
        \delta \Gamma^{\mu' \nu' \rho'}_{3\text{g}}
        D^{\mu \mu'}(K)
        D^{\nu \nu'}(P)
        D^{\rho \rho'}(R)
        \biggr\}.
\end{align}
Here, $R \equiv - K - P$, and the measure for a single momentum is defined as $[e^{\gamma_{\text{E}}} \Lh^{2} / (4 \pi)]^{\epsilon} \mathrm{d}^{d+1}\! K / (2 \pi)^{d+1}$, with $\gamma_{\text{E}}$ the Euler--Mascheroni constant. This expression contains the entirety of the $p_{-2}$ and $p_{-1}$ coefficients, and part of the constant $p_{0}$. To proceed further, one may perform the following two steps:
\begin{enumerate}
    \item Rescale the magnitudes of momenta by $|K| \mapsto \mE |K|$ etc., and introduce the hatted notation on the HTL functions. This extracts the $\mE$ dependence from all integrals.
    \item Change variables in the $K$ and $P$ integrals to write them as integrals over the magnitude of the four vectors $|K|$ and $|P|$ and the remaining angles: namely $(\Phi_K, \Phi_P, \theta)$ with ${\tan \Phi_K = |\vec{k}| / K_0}$, ${\tan \Phi_P = |\vec{p}| / P_0}$, and $\cos \theta = \hat{\vec k} \cdot \hat{\vec p}$. Here, the four-vectors are written as $K = (K_0, \vec{k})$ and $\hat{\vec k} \equiv \vec{k}/|\vec{k}|$. 
    \item Further transform from magnitudes of the momenta $(|K|,|P|)$ to Euclidean
     polar $(X,\chi)$ coordinates, given by $|K| = X \sin \chi$, $|P| = X \cos \chi$, with ${\chi \in [0, \pi / 2]}$, $X \in [0, \infty]$. 
\end{enumerate}
Here, we stress that by moving to the $(X, \chi)$ coordinates, one only changes the magnitudes $|K|, |P|$, and so $\hat{K}, \hat{P}$ remain unchanged.  After these steps, the integration measure becomes
\begin{equation}
    \int_{KP} = 
    C(d) \!
    \int_{\Omega}
    \int_{0}^{\pi / 2}\!\!\!\!  \ud \chi \, \sin^{d} \chi \cos^{d} \chi 
    \int_{0}^{\infty}\!\!\! \ud X \, X^{2d + 1},
\end{equation}
with the definitions:
\begin{equation}
%\begin{split}
%
    C(d) \equiv  \left (\frac{e^{\gamma_\text{E}}\Lh^2}{4\pi \mE^{2}}\right )^{3-d}\!\!\!\!\!\frac{4\pi^{d-\frac{1}{2}}}{(2\pi)^{2d+2}\Gamma\left (\frac{d}{2} \right )\Gamma\left (\frac{d-1}{2}\right )}; %\\
    \quad \int_{\Omega} \equiv  
    \prod_{i \in \{K, P\}}
    \int_{0}^{\pi} \!\!\ud \Phi_i \sin^{d-1} \Phi_i  
    %\!\int_{0}^{\pi}\!\!\! \ud \Phi_P \sin^{d-1} \Phi_P 
    \int_{0}^{\pi} \!\!\ud \theta \sin^{d-2} \theta.  
%
%\end{split}
\end{equation}
In all cases, the radial $X$ integral can be performed analytically, and in most of the terms of \eq\eqref{eq:ItotUV} the angular $\chi$ integral can also be performed analytically, even in general $d$. Physically, the radial $X$ integral generates one $1 / \epsilon$ divergence from the UV, and in terms with a $1 / \epsilon^2$ contribution, the $\chi$ integral generates the second, arising from the boundaries near $\chi = 0$ or $\pi / 2$. Translating this back to the original ($|K|$, $|P|$) coordinates, one sees these double divergences arise when $|K| \gg |P| \gg 0$ or $|P| \gg |K| \gg 0$, precisely matching the analysis conducted in \Ref\cite{Gorda:2018gpy}.  In the final term of \eq\eqref{eq:ItotUV}, containing $\delta \Gamma_{3\text{g}}$, the $\chi$ integral cannot be performed analytically, but it is finite. 

Before doing the $\chi$ integral, one can further simplify the tensor structure of the terms of the form $\hat{P} \cdot \DDN{n}(K) \cdot \hat{P}$ in \eq\eqref{eq:ItotUV} in general $d$. This procedure uses certain angular averages performed in \Ref\cite{Ee:2017}. This removes the products of four-momenta with the self-energies, leaving only traces of powers of the self energies. One may then perform the $\chi$ integration analytically, where possible, and expand near $d = 3$ to arrive at the coefficients $p_{-2}$ and $p_{-1}$ above, and part of the $p_{0}$ result. 
%, though there is the additional contribution from the original UV-finite pieces (to which we will return in a moment). 
Note that one must use the full $d$-dimensional self energies and three-gluon HTL vertex function in \eq\eqref{eq:ItotUV} to obtain the correct results \cite{Gorda:2021kme}. %These expressions are given in the detailed appendices of \Ref\cite{Gorda:2021kme}.

In the $p_{-2}$ coefficient, one finds that only the terms containing $\DDN{2}$ in \eq\eqref{eq:ItotUV} contribute: the terms containing only $\DDN{1}$ cancel each other, and the term containing $\delta \Gamma_{3\text{g}}$ only contributes to the $p_{-1}$ because the $\chi$ integral is finite. This leads to the dramatically compact expression for the integrand in \eq\eqref{eq:pm2_result} for $p_{-2}$.

The $p_{-1}$ result shown in \eq\eqref{eq:pm1_result} contains three terms, the first two of which are straightforward to compute. However, the final term contains the three-gluon HTL vertex function, which requires more work to evaluate.  In general, these HTL vertex functions are given by a $d$-dimensional integral representation. To reduce the numerical complexity of the expressions, one may introduce a modified Feynman parametrization, which allows one to perform these $d$-dimensional integrals order-by-order in $\epsilon$. This leaves only the one Feynman parameter left to be integrated over for each vertex. One finds that the contribution to $p_{-1}$ from this vertex function contains only a particularly simple contraction between Lorentz indices, since the propagators $D^{\mu \nu}(K) \mapsto \delta^{\mu \nu} / K^{2}$ in the UV (in Feynman gauge). However, the contractions contributing to $p_{0}$ are considerably more complicated, since the full longitudal and transverse structures within $D^{\mu \nu}(K)$ contribute. To deal with these structures, one can repeatedly use the generalized Ward identities to trade spatial contractions for temporal contractions and self-energy terms, both of which are easier to evaluate \cite{Gorda:2021kme}.  With this approach, the contributions to the final result from these vertex functions can be written as six-dimensional integrals (over $\chi$, $\Phi_K$, $\Phi_P$, $\theta$, and two Feynman parameters). These integrals can be computed numerically using Monte Carlo integration provided by the CUBA library \cite{Hahn:2004fe}. This provides one with the final contribution to the $p_{-1}$ in \eq\eqref{eq:pm1_result}, and a part of $p_{0}$. 

This completes the evaluation of the UV-sensitive terms in \eq\eqref{eq:ItotUV}, and leaves one with only the original UV-finite pieces to compute. These remaining finite pieces contain the full HTL-resummed propagators, as well as the three- and four-gluon HTL vertex functions, and therefore there are no general simplifications that arise from the contractions in these terms.
However, the finite nature of these terms makes them simpler to evaluate numerically, as one may directly set $d = 3$ in them. For this reason, they can be evaluation in an automated fashion, by performing any remaining non-Lorentz-invariant tensor contractions using an adapted version of the implementation discussed in \Ref\cite{Shtabovenko:2016sxi} (although in Euclidean space). Properties of the vertex corrections, in particular the generalized Ward identities and their tracelessness, are used extensively to make the resulting expressions as simple as possible. From the simplified expressions, one may perform the same steps 1-3 as presented above, and once again the $X$ integrals can be performed analytically. In the end, one must again perform up to six-dimensional numerical integrals to derive the remaining part of the coefficient $p_{0}$ presented in \eq\eqref{eq:p0_result}.  

\section{Results and Discussion}
\label{sec:results}

\begin{figure}[t]
% Use the relevant command for your figure-insertion program
% to insert the figure file.
\centering
\includegraphics[width=0.8\textwidth]{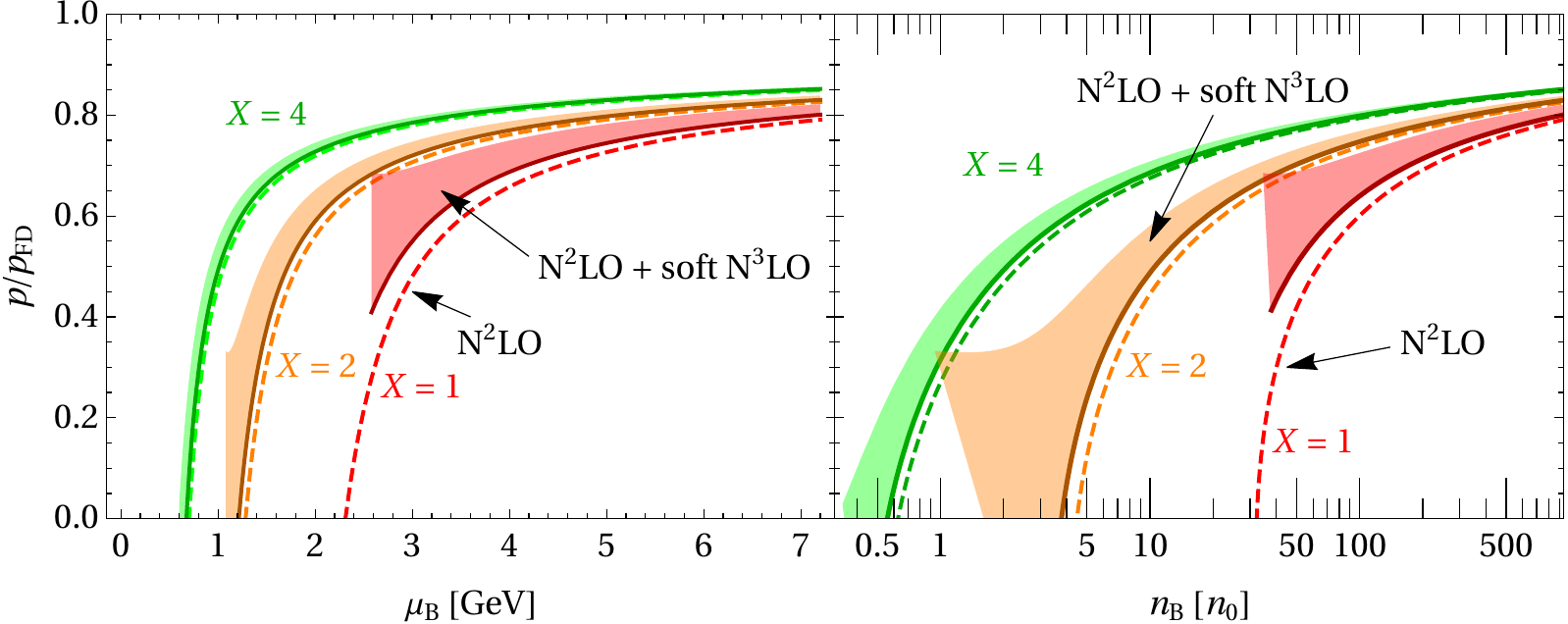}
    \caption{A comparison between the state-of-the-art pressure of cold QM at partial N3LO, including the interactions between soft screened modes, and the corresponding N2LO pressure given as functions of the baryon chemical potential $\muB$ (left) and baryon number density $\nB$ (right). Both pressures are shown normalized by the free Fermi--Dirac pressure $p_{\rm FD}$. The three colors correspond to the renormalization-scale variation in the hard sector, and the solid lines in the N3LO result correspond to fixing the factorization scale $\Lh$ to the PMS value. The filled bands in the N3LO result correspond to varying $\Lh$ by a factor of two in both directions about the PMS value.}
\label{fig:results}       % Give a unique label
\end{figure}

The analytic and numerical results for the soft contribution to the pressure of cold QM at N3LO were already shown above in Eqs.~\eqref{eq:pm2_result}-\eqref{eq:p0_result}; a plot of these results is shown in Fig.~\ref{fig:results}. The functional form of the result for the pressure at this order allows one to use the principle of minimal sensitivity to set the factorization scale $\Lh$ by solving ${\rm d}p / {\rm d} \Lh = 0$. One finds $\Lambda^\text{PMS}_{\text{h}} = \exp[-p_{-1}/(2p_{-2})]\mE$. In Fig.~\ref{fig:results}, using this value corresponds to the thick lines on the lower boundary of the shaded partial N3LO result. The result also still has a dependence on the UV renormalization scale $\bar{\Lambda}$, which we display over the usual range of $X \in \{1,2,4\}$ in different colors. We note here that the PMS value of $\Lh$ also corresponds to the one which maximizes the $\bar{\Lambda}$ dependence of the result: this means that the PMS value for $\Lh$ is the most conservative choice within this partial N3LO result on two fronts.

As can be seen from Fig.~\ref{fig:results}, including the contribution from the interacting soft screened modes at N3LO shifts the pressure to slightly higher values, and thus slightly decreases the UV-renormalization-scale dependence of the result; that is, including these interactions slightly improves the convergence of the pressure of cold QM. Note that this is contrary to the case at high $T$, where these soft modes limit the convergence of the pressure \cite{Blaizot:2003iq}. 

We note a further interesting observation, namely that the $p_{-2}$ coefficient seems to be related to the one-loop beta function of pure gauge theory, corresponding to the running of the $\mE$ parameter in the N2LO result. This connection has been investigated even further by the authors of \Ref\cite{Fernandez:2021jfr}, where an all-orders resummation for the leading and next-to-leading contributions from these soft modes has been conducted. 

Finally, we note that the improvements to the pressure presented above and in \Refs\cite{Gorda:2021kme,Gorda:2021znl} may have implications for the NS-matter EoS, and the improved convergence of the results is in agreement with analyses conducted using astrophysical observations to constrain the NS-matter EoS \cite{Annala:2019puf}. There it has been found (as also can be seen from Fig.~\ref{fig:renorm_scale_ep} above) that the NS-matter EoS follows the behavior of the pQCD cold-QM EoS to much lower densities than where pQCD calculations are currently converged. This result certainly motivates further improvements to the cold-QM EoS of pQCD matter, which are currently underway.

\section*{Acknowledgments}

This proceedings is a summary of work conducted in collaboration with Aleksi Kurkela, Risto Paatelainen, Saga Säppi, and Aleksi Vuorinen. TG was in part supported by the Deutsche Forschungsgemeinschaft (DFG, German Research Foundation) -- Project-Id 279384907 -- SFB 1245.

%% For two-column wide figures use syntax of figure~\ref{fig-2}
%% \begin{figure*}
%% \centering
%% % Use the relevant command for your figure-insertion program
%% % to insert the figure file. See example above.
%% % If not, use
%% \vspace*{5cm}       % Give the correct figure height in cm
%% \caption{Please write your figure caption here}
%% \label{fig-2}       % Give a unique label
%% \end{figure*}

%%For figure with sidecaption legend use syntax of figure
%%\begin{figure}
%%% Use the relevant command for your figure-insertion program
%%% to insert the figure file.
%%\centering
%%\sidecaption
%%%\includegraphics[width=5cm,clip]{none.pdf}
%%\caption{Please write your figure caption here}
%%\label{fig-3}       % Give a unique label
%%\end{figure}

%% For tables use syntax in table~\ref{tab-1}.
%% \begin{table}
%% \centering
%% \caption{Please write your table caption here}
%% \label{tab-1}       % Give a unique label
%% % For LaTeX tables you can use
%% \begin{tabular}{lll}
%% \hline
%% first & second & third  \\\hline
%% number & number & number \\
%% number & number & number \\\hline
%% \end{tabular}
%% % Or use
%% \vspace*{5cm}  % with the correct table height
%% \end{table}
%
% BibTeX or Biber users please use (the style is already called in the class, ensure that the "woc.bst" style is in your local directory)
\bibliography{refs.bib}

\end{document}